\documentclass[aps,prx,twocolumn,floats,showpacs,superscriptaddress,nofootinbib,longbibliography]{revtex4-2}
\usepackage{graphicx,epsfig}
\usepackage{times,bbm}
\usepackage{graphics,dcolumn,bm,float}
\usepackage{amssymb,amsmath,rotate,color,amsfonts}
\usepackage[title,titletoc,toc]{appendix}
\usepackage{mathtools}
\usepackage{booktabs}
\usepackage{enumitem}
\usepackage{tcolorbox}
\usepackage[pagebackref=false,colorlinks,linkcolor=magenta,citecolor=blue,urlcolor=magenta]{hyperref}

\usepackage{wrapfig}
\usepackage{lipsum}
\usepackage{mwe}
\usepackage[mathlines]{lineno}
\usepackage[mathscr]{euscript}
\usepackage{hyperref}
\usepackage{breqn}
\usepackage{bbold}
\usepackage{pgfplots}
\usepackage{float}
\usepackage{tkz-euclide}
\usepackage{braket}
\usepackage{physics}
\usepackage[caption=false]{subfig}
\usepackage[export]{adjustbox}
\usepackage{tikz}
\usepackage{slashed}
\usepackage{bm}
\usetikzlibrary{through,calc}
\usetikzlibrary{positioning}

\newcommand{\be}{\begin{equation}}
\newcommand{\ee}{\end{equation}}
\newcommand{\ba}{\begin{align}}
\newcommand{\ea}{\end{align}}

\newcommand{\bea}{\begin{eqnarray}}
\newcommand{\eea}{\end{eqnarray}}
\newcommand{\nn}{\nonumber}

\newcommand{\bmt}{\left[\begin{matrix}}
\newcommand{\emt}{\end{matrix}\right]}

\begin{document}
\preprint{}
\title{Zeeman-like coupling to valley degree of freedom in Si-based spin qubits}

\author{S. A. Jafari}
\affiliation{2nd Institute of Physics C$,$ RWTH Aachen University$,$ 52074 Aachen$,$ Germany}
\email{akbar.jafari@rwth-aachen.de}
\author{Hendrik Bluhm}
\affiliation{2nd Institute of Physics C$,$ RWTH Aachen University$,$ 52074 Aachen$,$ Germany}
\author{David P. DiVincenzo}
\affiliation{Institute for Quantum Information$,$ RWTH Aachen University$,$ D-52056 Aachen$,$ Germany}
\affiliation{Peter Gr\"unberg Institute$,$ Theoretical Nanoelectronics$,$ Forschungszentrum J\"ulich$,$ D-52425 J\"ulich$,$ Germany}
\date{\today}

\begin{abstract}
Increasing the valley splitting in Si-based heterostructures is critical for improving the performance of semiconductor qubits. This paper demonstrates that the two low-energy conduction band valleys are not independent parabolic bands. Instead, they originate from the X-point of the Brillouin zone, where they are interconnected by a degeneracy protected by the non-symmorphic symmetry of the diamond lattice. This semi-Dirac-node degeneracy gives rise to the $\Delta_1$ and $\Delta_{2'}$ bands, which constitute the valley degrees of freedom. By explicitly computing the two-component Bloch functions $X_1^\pm$, using the wave vector group at the X-point, we determine the transformation properties of the object $(X_1^+,X_1^-)$. We demonstrate that these properties are fundamentally different from those of a spinor. Consequently, we introduce the term "valleyor" to emphasize this fundamental distinction. The transformation properties of valleyors induce corresponding transformations of the Pauli matrices $\tau_1,\tau_2$ and $\tau_3$ in the valley space. Determining these transformations allows us to classify possible external perturbations that couple to each valley Pauli matrix, thereby identifying candidates for valley-magnetic fields, $\vec {\cal B}$. These fields are defined by a Zeeman-like coupling $\vec{\cal B}\cdot\vec\tau$ to the valley degree of freedom. In this way, we identify scenarios where an applied magnetic field $\vec B$ can leverage other background fields, such as strain, to generate a valley-magnetic field $\vec{\cal B}$. 
This analysis suggests that beyond the well-known mechanism of potential scattering from Ge impurities, there exist additional channels—mediated by combinations of magnetic and strain-induced vector potentials—to control the valley degree of freedom.   
\end{abstract}



\maketitle
\narrowtext

\section{Introduction}
\begin{figure}[bt]
\centering
\includegraphics[width=3.5cm,angle=0]{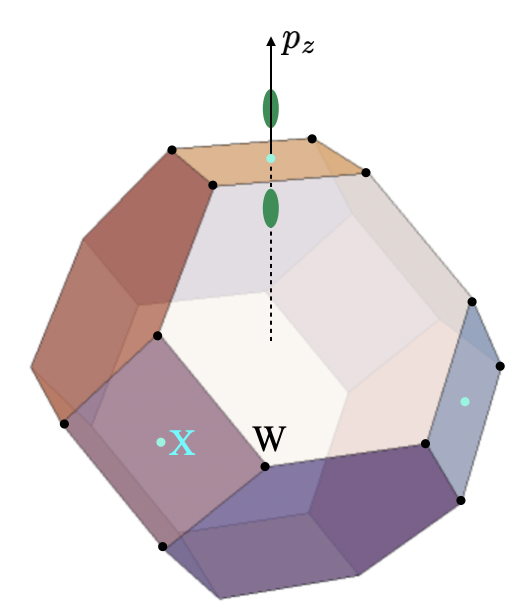}
\includegraphics[width=4.1cm]{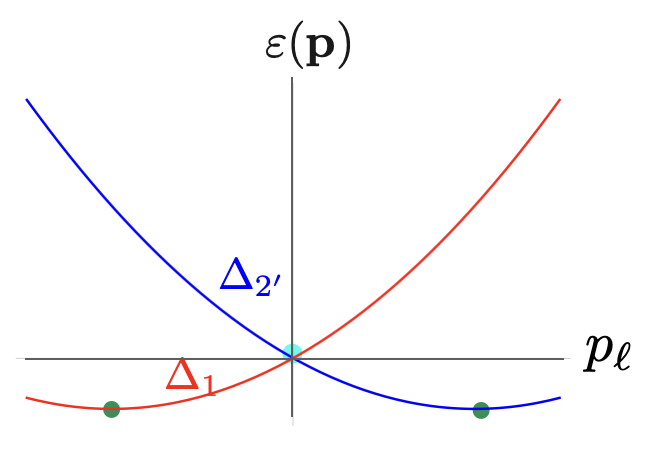}
\caption{
(Left) Brillouin zone of the face-centered cubic lattice. The $X$ point (cyan) lies at the center of a square face, with four equivalent $W$ points indicated by black dots around each $X$ point. The two Si valleys (green ellipsoids) are located at $\pm k_1$ from the $X$ point along the longitudinal ($z$) direction. (Right) Energy dispersion of the $\Delta_1$ (red) and $\Delta_{2'}$ (blue) bands near the $X$ point in Si, shown as a function of longitudinal momentum $p_\ell = p_z$. These bands, which form the Si valleys, emerge from the two-dimensional representation of the space group at the $X$ point when moving away from this high-symmetry location. The energy of the $X$ point is about $80$ meV above the degenerate conduction band minimum.
}
\label{bands.fig}
\end{figure}

Bulk Si features six equivalent conduction band minima, commonly referred to as valleys~\cite{Cardona2010,Coopersmith2013}. In Si/SiGe heterostructures grown along the $z$-direction, strain lifts the degeneracy, causing the four transverse valleys to shift to higher energies. This leaves two degenerate valleys in the longitudinal ($z$) direction, denoted by the states $|\pm\rangle$, as the low-energy degrees of freedom. These two valleys are situated on the $\Delta$-line connecting the Brillouin zone center ($\Gamma$) to the center of the square face ($X$), as depicted in Fig.~\ref{bands.fig}. Specifically, the valleys are located at wavevectors $\pm k_0=\pm 0.85\times 2\pi/a$ from the $\Gamma$ point, or equivalently, at $\pm k_1=0.15\times 2\pi/a$ from the six equivalent $X$ points, where $a=5.43\AA$ is the lattice constant of Si~\cite{Friesen2007}.

Spin qubits in such heterostructures are promising candidates for semiconductor-based quantum computing. These qubits utilize the spin degree of freedom of electrons confined in a soft electrostatic potential, $V_{\rm Dot}(x,y)$. A key challenge in achieving reproducible qubit properties is the near-degeneracy of the valley states, $|\pm\rangle$. This degeneracy can spoil the qubit's Hilbert space by coupling it to the valley degree of freedom. Although sufficiently large valley splittings have been measured on average \cite{philips_universal_2022,marcks_valley_2025}, recent studies indicate that the splitting is dominated by alloy disorder \cite{paquelet_wuetz_atomic_2022}, leading to a random distribution with a significant probability of near-zero values, as confirmed experimentally \cite{volmer_mapping_2024}. Proposed strategies to avoid these undesirably small splittings include realizing sharper interfaces \cite{paquelet_wuetz_atomic_2022,Lima2023} or introducing an oscillatory Ge concentration \cite{McJunkin2022}; however, these approaches remain experimentally challenging \cite{gradwohl_enhanced_2025}. A combination of shear strain and concentration oscillations with a longer period is predicted to enhance both the valley splitting and spin-orbit coupling \cite{woods_spin-orbit_2023,woods_coupling_2024}. Therefore, designing alternative mechanisms to achieve a more stable and controllable valley splitting is highly desirable.

Similar to spin states $|\uparrow\rangle$ and $|\downarrow\rangle$, it is tempting to treat the two valley states $|\pm\rangle$ as a two-component spinor and define corresponding Pauli matrices $\tau_j$ (for $j=1,2,3$) analogous to the spin Pauli matrices $\sigma_j$. However, it is clear from the outset that the valley degree of freedom cannot transform like spin. If the transformation properties of $\sigma$ and $\tau$ under crystal rotations were identical, one would expect a Zeeman-like coupling $\propto \vec B \cdot \vec \tau$ analogous to the Zeeman coupling $\mu_B \vec B \cdot \vec\sigma$. Such a term would naturally split the valley states, just as a magnetic field splits spin states—a phenomenon that is not observed. The purpose of this paper is to identify the valley-magnetic fields $\vec {\cal B}$ that couple to the valley degree of freedom via $\vec\tau$, and to express these fields in terms of more fundamental physical fields.

The dominant approach to designing valley-splitting mechanisms relies on the fact that the two valley minima at momenta $\pm \vec q$ can be mixed by any "potential scattering" process that provides a Fourier component at $2\vec q$, where $\vec q$ can be either $\vec k_0$ or $\vec k_1$~\cite{Kantner2024}. This logic, which applies to any pair of valley minima in the Brillouin zone, treats them simply as two unrelated points in momentum space separated by a wave vector $2\vec q$. However, it does not utilize the crucial fact that, in Si, these minima originate from a two-fold degeneracy at the high-symmetry $X$ point, as shown in Fig.~\ref{bands.fig}.

Given the importance of Si for quantum technologies, it is timely to address a fundamental question regarding the very nature of the valley degree of freedom and to derive the transformation properties of the valley Pauli matrices $\vec\tau$. This is necessary to classify all symmetry-permitted ways in which external perturbations can couple to these degrees of freedom. A key message of this paper is that \emph{the valley degrees of freedom do not exhibit spinor-like behavior}; their transformation properties under space group operations are fundamentally different. Therefore, referring to them as a 'pseudospin' can be misleading, as it implies a transformation rule similar to that of spinors. To emphasize this distinction, we introduce the term valleyor to describe the two-component object in the space spanned by the two Bloch states $ |\pm\rangle \leftrightarrow X_1^\pm $.

The purpose of this paper is to employ the non-symmorphic wave vector group at the $X$ point of the Brillouin zone of bulk Si to compute the explicit form of the valleyor basis states $X_1^\pm$ and determine their transformation properties. This approach yields the following key results:
(1) We identify quadratic terms in the effective Hamiltonian that are as significant as the parabolic band components. This finding corroborates a recent tight-binding calculation~\cite{Chamon2020} and associates the resulting nodal line structure—located slightly above the conduction band minima ($\sim 80$ meV)—with the valley degrees of freedom.
(2) We provide a classification of how various vector and tensor perturbations (e.g., strain) can couple to the valley degrees of freedom. This extends the concept beyond the prevailing mechanism, where scattering by an external scalar potential is considered the primary source of valley coupling~\cite{Friesen2007}. 
We show how a magnetic field $\vec B$ can be leveraged, in combination with other background fields, to produce a valley-magnetic field $\vec{\cal B}$. 

This paper is organized as follows:
Section~\ref{secmotivation} combines the two oppositely located valleys and discusses the ambiguity in assigning the valley Pauli matrices, demonstrating how symmetry considerations resolve this issue. Section~\ref{secgroupX} provides a pedagogical presentation of the wave vector group at the $X$ point. This section is intentionally detailed, as information on the space group at $X$ is not readily accessible in the literature\footnote{Among the classic textbooks, Jones~\cite{Jones1959} contains an error in the character table, while Yu and Cardona~\cite{Cardona2010} present notational inconsistencies. Furthermore, the basis functions provided by Jones~\cite{Jones1959} do not correctly reduce to the $\Delta_1$ and $\Delta_{2'}$ bands in Fig.~\ref{bands.fig}, which are the proper labels for the valley degrees of freedom.}. In Section~\ref{secBasisTransform}, we compute the transformation properties of the valley Pauli matrices to determine their symmetry type, i.e., we identify the irreducible representations to which they belong. We also classify various other vector, axial vector, and tensorial combinations according to the irreducible representations of the wave vector group at the $X$ point. Finally, Section~\ref{secCoupling} synthesizes the information from Section~\ref{secBasisTransform} to construct all permissible coupling forms to the valley Pauli matrices $\tau_1,\tau_2,\tau_3$ in the form of $\vec{\cal B}\cdot\vec{\tau}$, summarizing them in a table for future reference.

\section{A new effective model for silicon valleys}
\label{secmotivation}
The parabolic dispersion near the bottom of the conduction bands in bulk Si is given by two parabolas 
\begin{equation}
   \frac{p_z^2\pm 2p_z \hbar k_1}{2m_l}+\frac{p_x^2+p_y^2}{2m_t},
   \label{oldSi.eqn}
   \end{equation}
with the two minima (valleys) located at $\mp k_1$ with respect to the $X$ point where $k_1\approx 0.15\times 2\pi/a$, and the energy of the bottom of the conduction bands is $-\hbar^2k_1^2/(2m_l)$.  
In Si/SiGe quantum wells, the three pairs of valleys are split, leaving one pair along the growth ($z$) direction as the lowest-energy states, with the two in-plane pairs shifting to higher energies. A key requirement for Si-based spin qubits is to lift the degeneracy of these two lowest valleys with a valley splitting energy $E_{\rm vs}$. This ensures that qubit gate operations, performed at energy scales below $E_{\rm vs}$, address a purely spin-based qubit Hilbert space. Existing models for coupling between the two valleys are based on treating them as independent minima in momentum space, located at $\pm \vec k_0$ from the Brillouin zone center $\Gamma$, where $k_0=0.85\times2\pi/a$~\cite{Kantner2024}.

Denoting the ground state wave function of the parabolic band
\begin{equation}
\frac{p_z^2}{2m_l}+\frac{p_x^2+p_y^2}{2m_t}
\end{equation}
by $\phi_0$, the ground states $\phi_\pm$ of the two shifted parabolas in Eq.~\eqref{oldSi.eqn} can be obtained by a simple gauge transformation as
\begin{equation}
\phi_\pm(z)=\langle z|\pm\rangle=e^{\mp i k_1 z}\phi_0(z).
\label{shift.eqn}
\end{equation}
The valley splitting is then naturally given by an off-diagonal matrix element $V_{-+}$, which becomes~\cite{Friesen2007,Friesen2010,Hosseinkhani2020,McJunkin2022,Losert2023,Lima2023}
\begin{equation}
\langle -|V|+\rangle=\int e^{2ik_1 z}|\phi_0(z)|^2 V(z) dz.
\label{VS.eqn}
\end{equation}
A similar analysis for valleys located at $\pm \vec k_0$ within the first Brillouin zone yields an analogous formula with $k_1$ replaced by $k_0$. Both resonances are permissible sources of valley splitting within this scattering formalism~\cite{Kantner2024}. The central question addressed in this paper is whether there exist other valley splitting mechanisms beyond this scattering matrix element formulation.

To appreciate the need for transformation properties of valley Pauli matrices, we begin by noting that the naive incorporation of the valley degree of freedom into Si-based spin qubits ignores the transformation properties of the $\tau_j$ matrices\footnote{We deliberately label the valley matrices with $\tau_1,\tau_2,\tau_3$ and \emph{not} with $\tau_x,\tau_y,\tau_z$ to avoid associating these Pauli matrices with the spatial directions $x,y,z$. This distinction will become clear in the subsequent derivations.}. Indeed, treating the two parabolas as degenerate sectors of a two-valley theory without imposing constraints from symmetry operations yields the Hamiltonian
\begin{equation}
\left(\frac{p_z^2}{2m_l}+\frac{p_x^2+p_y^2}{2m_t}\right)\tau_0+v_1 p_z\tau_j,
\end{equation}
where $\tau_0$ is the $2\times 2$ identity matrix and the velocity $v_1=\hbar k_1/m_l=2.04\times 10^5$ m/s defines the separation of the two Si valleys from the $X$ point. At this naive level, where $\tau_j$ must simply yield an eigenvalue of $\pm1$ to reproduce Eq.~\eqref{oldSi.eqn}, it can be any of the Pauli matrices $\tau_1,\tau_2,\tau_3$. For any such choice, the above matrix Hamiltonian reduces to the two parabolas in Eq.~\eqref{oldSi.eqn} and yields the same eigenfunctions~\eqref{shift.eqn} for the shifted valleys. 

It is well established from the early days of quantum theory of solids that the conduction band minima in Si belong to the $X_1$ representation of the wave vector group at the $X$-point~\cite{Lax1961,Hensel1965,Cardona2010}. Therefore, to construct a faithful representation of the valley degree of freedom, the explicit form of the two basis functions forming this representation must be considered. These two Bloch orbitals, denoted $X_1^+$ and $X_1^-$, are degenerate at the $X$ point. Upon moving away from the $X$ point along the $\Delta$-line (see Fig.~\ref{bands.fig}), they split into the $\Delta_1$ and $\Delta_{2'}$ bands, which ultimately form the conduction band minima along the $z$-direction.

One possible form for these basis functions $\tilde X^+_1$ and $\tilde X^-_1$ is given in Ref.~\cite{Jones1959}. However, these do not reduce to the $\Delta_1$ and $\Delta_{2'}$ bands when moving away from the $X$ point. To correctly label the valley degrees of freedom, they must be linearly combined to ensure compatibility with these bands. To find the appropriate basis functions $X^+_1$ and $X^-_1$ satisfying this requirement, we compute them explicitly by projecting free electron bands at the $X$ point onto the $X_1$ representation. The use of a free electron basis does not affect the symmetry properties for two reasons: (i) Only the correct transformation character under the wave vector group operations is required. The complexity of the wave function is irrelevant; a simple free-electron wave function with the correct characters suffices. (ii) The degeneracy at the $X$ point is protected by non-symmorphic symmetry\footnote{This protection extends along the entire $Z$ line connecting the $X$ and $W$ points in the left panel of Fig.~\ref{bands.fig}.}, meaning that including interactions will not split it and thus does not alter the qualitative features of the free electron bands near the $X$ point.

For the sake of completeness, we provide a detailed and pedagogical account of the computations, as the wave vector group at the $X$ point of the Si structure~\cite{Herring1942,Lax1961} is a non-symmorphic example not easily accessible in standard textbooks~\cite{Jones1959,Cardona2010}. Obtaining the explicit form of the basis functions for the $X_1$ representation at the $X$ point allows us to compute the transformation properties of the valley degrees of freedom $\tau_j$ under the symmetry operations of this group. We compute the multiplication table for the irreducible representations of the space group at the $X$ point, which is essential for classifying possible perturbations that can couple to the valley Pauli matrices $\tau_1$, $\tau_2$, and $\tau_3$. This analysis reveals possible mechanisms for valley splitting that operate beyond scalar potential scattering, providing an alternative resources/strategies for valley splitting. 

\section{Group of wave vector at $X$-point}
\label{secgroupX}
This section provides a self-contained account of the wave vector group at the $X$-point. Readers willing to accept the explicit form of the valley Bloch functions in Eq.~\eqref{mybasis.eqn} may skip this section. As noted, the detailed structure of this group for the diamond lattice is available in only a few sources~\cite{Herring1942,Jones1959}. For instance, Ref.~\cite{Herring1942} provides the character table but not the basis functions, which are computed in Ref.~\cite{Jones1959} using an old-fashioned notation that is difficult to follow without consulting other parts of the book. Parts of Jones's calculations are reproduced in Ref.~\cite{Cardona2010} with modern notation. While the presentation in Ref.~\cite{Cardona2010} corrects a typo in the character table given by Jones~\cite{Jones1959}, it introduces another error by using the mirror operators $\sigma_x: xyz\to \bar xyz$ and $\sigma_y:xyz\to x\bar y z$ in place of the correct diagonal mirrors $m_1:xyz\to yxz$ and $m_2:xyz\to \bar y\bar x z$. This leads to inconsistencies in determining the basis functions for the classes of the space group at the $X$-point. Therefore, we present a self-contained review with corrections, computing the basis functions and the products of irreducible representations. We follow the notation of Ref.~\cite{Cardona2010}, which is consistent with that of Ref.~\cite{DresselhauseGroupTheory}.

The point group of the $X$-point in the Si structure, which must be a subgroup of the tetrahedral group $T_d$, is $D_{2d}$. This group is specified by the following elements:
\begin{eqnarray}
   D_{2d}~~&&= \{E,C_{4z}^2,C_{4x}^2,C_{4y}^2,2S_{4z},m_1,m_2\},
   \label{D2dops.eqn}\\
   &&\equiv \{xyz,\bar x\bar y z,x\bar y\bar z, \bar xy\bar z,\bar yx\bar z,y\bar x\bar z,yxz,\bar y\bar x z \},
   \label{pointgroup.eqn}
\end{eqnarray}
where the second line represents the effect of the operators in the first line on the coordinates $(x,y,z)$ compactly represented as $xyz$, using the notation $\bar x=-x$, etc. The essential non-symmorphic aspect of the diamond structure's space group is the presence of a diagonal quarter translation $T$ by the vector $\frac{a}{4}(1,1,1)$, which connects the two identical atoms in the Si diamond lattice. The glide operator of the non-symmorphic diamond lattice is constructed as $g=Tm_z$, where $m_z: xyz \to xy\bar z$ is the mirror reflection with respect to the $z=0$ plane. Including this glide operator generates 8 additional symmetry operators given by
\begin{eqnarray}
   g\times\{E,C_{4z}^2,C_{4x}^2,C_{4y}^2,2S_{4z},m_1,m_2\}.
\end{eqnarray}
However, the above set is not yet closed. Since $D_{2d}$ itself is a group, the resulting $16$ elements would only form a larger group if the glide operator $g$ commuted with all elements of $D_{2d}$. The key to identifying the missing elements lies in quantifying this non-commutativity. For this purpose, let us compare, for example, $C^2_{4x}g$ and $gC^2_{4x}$. Their effects on the coordinates $(x,y,z)$ are as follows~\cite{Cardona2010}:
\begin{eqnarray}
   &&C^2_{4x}g(x,y,z)=C^2_{4x}T(x,y,\bar z),\nn\\
   &=&C^2_{4x}(x+a/4,y+a/4,-z+a/4),\nn\\
   &=&(x+a/4,-y-a/4,z-a/4),
\end{eqnarray}
whereas,
\begin{eqnarray}
&&gC^2_{4x}(x,y,z)=g(x,\bar y,\bar z)=T(x,\bar y,z),\nn\\
&=&(x+a/4,-y+a/4,z+a/4).
\end{eqnarray}
Comparing the final results of these two equations suggests introducing a face diagonal translation
\begin{equation}
Q=(0,\bar a/2,\bar a/2),
\label{Q.eqn}
\end{equation}
which allows us to precisely quantify the non-commutativity of $g$ and $C^2_{4x}$ as
\begin{equation}
C^2_{4x}g=QgC^2_{4x}.
\end{equation}
Therefore, in addition to the glide operation $g$, the translation operator $Q$ must be included, along with their product $Qg$. This procedure yields a total of 32 elements. If we consider $C^2_{4y}$ in place of $C^2_{4x}$ in relation~\eqref{Q.eqn}, we would obtain $Q'=(\bar a/2,0,\bar a/2)$ instead of $Q$. Consequently, the resulting set, or {\em complex}, still does not appear to be closed. This is where the special role of the $X$-point becomes important. The coordinate of the $X$-point in $k$-space is $\vec k_X=\frac{2\pi}{a}(0,0,1)$. When the above set of 32 operators acts on a Bloch function of the form
\begin{equation}
\psi=e^{i\vec k_X\cdot\vec r}u(\vec r)=e^{i2\pi z/a}u(\vec r),
\end{equation}
where $u$ is the lattice-periodic part of the wave function, the complex becomes closed, forming the group of the wave vector at the $X$-point of the diamond structure. This is why the group of the wave vector at the $X$-point is sometimes referred to as a complex.

This extended group, like any other group, can be decomposed into conjugacy classes as follows~\cite{Jones1959,Cardona2010}:
\begin{eqnarray}
   {\cal C}_1&=&\{E \},\nn\\
   {\cal C}_2&=&\{C^2_{4x},C^2_{4y},QC^2_{4x},QC^2_{4y} \},\nn\\
   {\cal C}_3&=&\{C^2_{4z} \},\nn\\
   {\cal C}_4&=&\{Qgm_1,gm_2 \},\nn\\
   {\cal C}_5&=&\{gS_{4z}, gS_{4z}^{-1},QgS_{4z},QgS_{4z}^{-1}\},\nn
\end{eqnarray}
\begin{eqnarray}
   {\cal C}_6&=&\{g,Qg \}\nn,\\
   {\cal C}_7&=&\{gC^2_{4x}, gC^2_{4y},QgC^2_{4x},QgC^2_{4y}\},\nn\\
   {\cal C}_8&=&\{gC^2_{4z},QgC^2_{4z}\},\label{C114.eqn}\\
   {\cal C}_9&=&\{m_1,m_2 \},\nn\\
   {\cal C}_{10}&=&\{S_{4z}, S_{4z}^{-1},QS_{4z},QS_{4z}^{-1}\},\nn
\end{eqnarray}
\begin{eqnarray}
   {\cal C}_{11}&=&\{Qm_1,Qm_2 \},\nn\\
   {\cal C}_{12}&=&\{Qgm_2,gm_1 \},\nn\\
   {\cal C}_{13}&=&\{QC^2_{4z} \},\nn\\
   {\cal C}_{14}&=&\{Q \}.\nn
\end{eqnarray}
Note that we have corrected a notational confusion in Ref.~\cite{Cardona2010} by replacing $\sigma_x$ and $\sigma_y$ with the proper diagonal mirror reflections $m_1$ and $m_2$ given in Eqs.~\eqref{D2dops.eqn} and~\eqref{pointgroup.eqn}.

\begin{table}[tbp]
    \centering
    \begin{tabular}{|c|ccccc||ccccc||cccc|}
        \hline
	   &${\cal C}_1$ &${\cal C}_2$ &${\cal C}_3$ &${\cal C}_{4}$ &${\cal C}_{5}$ &${\cal C}_6$ &${\cal C}_7$ &${\cal C}_8$ &${\cal C}_{9}$ &${\cal C}_{10}$ &${\cal C}_{11}$ &${\cal C}_{12}$ &${\cal C}_{13}$ &${\cal C}_{14}$\\
        \hline
        $M_1$ & 1 & 1 & 1 & 1 & 1 & 1 & 1 & 1 & 1 & 1 & 1 & 1 & 1 & 1 \\ \hline
        $M_2$ & 1 & 1 & 1 & -1 & -1 & 1 & 1 & 1 & -1 & -1 & -1 & -1 & 1 & 1 \\ \hline
        $M_3$ & 1 & -1 & 1 & -1 & 1 & 1 & -1 & 1 & -1 & 1 & -1 & -1 & 1 & 1 \\ \hline
        $M_4$ & 1 & -1 & 1 & 1 & -1 & 1 & -1 & 1 & 1 & -1 & 1 & 1 & 1 & 1 \\ \hline
        $M_5$ & 2 & 0 & -2 & 0 & 0 & 2 & 0 & -2 & 0 & 0 & 0 & 0 & -2 & 2 \\ \hline
	\hline
	$M'_1$ & 1 & 1 & 1 & 1 & 1  & -1 & -1 & -1 & -1 & -1 & -1 & 1 & 1 & 1 \\ \hline
	$M'_2$ & 1 & 1 & 1 & -1 & -1 & -1 & -1 & -1 & 1 & 1 & 1 & -1 & 1 & 1 \\ \hline
	$M'_3$ & 1 & -1 & 1 & -1 & 1 & -1 & 1 & -1 & 1 & -1 & 1 & -1 & 1 & 1 \\ \hline
	$M'_4$ & 1 & -1 & 1 & 1 & -1 & -1 & 1 & -1 & -1 & 1 & -1 & 1 & 1 & 1 \\ \hline
        $M'_5$ & 2 & 0 & -2 & 0 & 0 & -2 & 0 & 2 & 0 & 0 & 0 & 0 & -2 & 2 \\ \hline
	\hline
	$X_1$ & 2 & 0 & 2 & 0 & 0 & 0 & 0 & 0 & 2 & 0 & -2 & 0 & -2 & -2 \\ \hline
	$X_2$& 2 & 0 & 2 & 0 & 0 & 0 & 0 & 0 & -2 & 0 & 2 & 0 & -2 & -2 \\ \hline
	$X_3$ & 2 & 0 & -2 & 2 & 0 & 0 & 0 & 0 & 0 & 0 & 0 & -2 & 2 & -2 \\ \hline
	$X_4$ & 2 & 0 & -2 & -2 & 0 & 0 & 0 & 0 & 0 & 0 & 0 & 2 & 2 & -2 \\ \hline
    \end{tabular}
    \caption{Irreducible representations of the space group at the X point of Diamond structure~\cite{Cardona2010}. This table 
    from Ref.~\cite{Cardona2010} has a correction in the last row with respect to the one in Ref.~\cite{Jones1959}.}
    \label{XIRtab}
\end{table}
The irreducible representations of the above extended group are provided in Table~\ref{XIRtab}~\cite{Cardona2010}.
This table has been deliberately partitioned into $M$-type, $M'$-type, and $X$-type representations. The block of $M$-type representations is isomorphic to the $D_{4} \equiv C_{4v}$ group. In fact, the first block of classes defines the $C_{4v}$ family of operations—which corresponds to the symmetry of the $\Delta$ line connecting the $\Gamma$ and $X$ points—as follows (cf. Table 10.9 of Ref.~\cite{DresselhauseGroupTheory}):
\begin{eqnarray}
   && {\cal C}_1 \to \{E|0\},\nn\\
   && {\cal C}_2 \to 2\{iC_4^2|T\},\nn\\
   && {\cal C}_3 \to \{C_{4}^2|0\},\label{C4vMap.eqn}\\
   && {\cal C}_4 \to 2\{iC'_2|0\},\nn\\
   && {\cal C}_5 \to 2\{C_4|T\},\nn
\end{eqnarray}
where $T$ is the diagonal translation defined below Eq.~\eqref{pointgroup.eqn}. The operations without $T$ define the abstract $C_{4v}$ point group. In the diamond lattice structure, the diagonal translation $T$ is required to map the lattice to itself. Even after including $T$ to form the symmetry operators of the diamond space group, the multiplication table remains isomorphic to the $C_{4v}$ point group. As can be seen, the $M'$-type representations are obtained by taking the direct product of the $M$-block with the two-element group $\{E,m_z\}$, which has the following representation:
\begin{table}[h]
   \begin{tabular}{c|cc}
         & $E$  & $m_z$\\
      \hline
         $M$ & $1$  & $1$\\
	 $M'$ & $1$  & $-1$
   \end{tabular}
\end{table}

For example, starting from the fact that the representation $M_1$ defines scalars, this direct product structure implies that the representation $M'_1$ is the pseudoscalar representation that changes sign under mirror ($m_z$) reflection. This pattern extends to all other representations in the $M$ and $M'$ families. The fact that the element $g$ contains $m_z$ (when translations are ignored) reveals that the first 10 classes possess a $D_{4h}=C_{4v}\otimes\{E,m_z\}$ structure. This structure is also evident in the grouping of classes in Eq.~\eqref{C114.eqn} and holds for the "direction group"—the group resulting from ignoring the translation parts of a space group.

However, within the space group that includes translations by $T$ and $Q$, the $M$ and $M'$ representations of the extended group at the $X$-point do not yield physically acceptable representations for the Bloch wave functions at the $X$-point—although they do provide acceptable representations for, e.g., the $\Delta$ line under the identification given in Eq.~\eqref{C4vMap.eqn}. The reason only the $X$-type representations are relevant for Bloch states at the $X$-point is that they satisfy
\begin{equation}
   Q\psi=e^{i2\pi(z-a/2)/a}u(x,y-a/2,z-a/2)=-\psi,
\end{equation}
where the periodicity of $u$ under translation by $Q$—a translation vector of the face-centered cubic lattice—has been used. The above equation implies that physically acceptable representations for the wave vector group, corresponding to permissible Bloch wave functions at the $X$-point, must have negative characters under $Q$. This condition is satisfied only by the $X$-type representations, which are all two-dimensional. Therefore, all physically acceptable Bloch wave representations at the $X$-point are two-fold degenerate. Since the $M$ and $M'$ representations have positive character under the $Q$ operation, Bloch wave functions cannot transform as $M$ or $M'$ types. However, other objects (as we will see) can possess $M$ and $M'$ symmetry types.

\subsection{Basis functions for $X_1$ representation}
The conduction bands in Si belong to the $X_1$ representation~\cite{Cardona2010}. The fact that the structure factor vanishes for the wave vector $2\vec k_X=4\pi(0,0,1)/a$ connecting two opposite $X$-points under the displacement $\vec T=a(1,1,1)/4$,
\begin{equation}
1+e^{i 2\vec k_X \cdot \vec T}=0,
\end{equation}
implies that no Bragg reflections connect $\vec k_X$ and $-\vec k_X$, leaving the $X$-point doubly degenerate. To compute the two degenerate basis functions that give rise to the two valleys, we follow~\cite{Jones1959} and form the lattice-periodic functions
\begin{equation}
   u_{\vec l}(\vec r)=\exp\left[-i\sum_{\alpha,\beta}l_\alpha G_{\alpha\beta}x_\beta\right],~~~
   \label{freeBloch.eqn}
\end{equation}
where $l_\alpha$ are integer, $(x_1,x_2,x_3)=(x,y,z)$ and the matrix $G_{\alpha\beta}$ defines the reciprocal lattice vectors
in the FCC lattice 
\begin{equation}
   G=\frac{2\pi}{a}\begin{bmatrix}
      -1 & 1 & 1 \\
      1  & -1 & 1\\
      1 & 1 & -1
   \end{bmatrix}.
   \label{reciprocal.eqn}
\end{equation}
For simplicity, in the following calculations we choose the unit of length such that $a=2\pi$, causing all factors of $a/2\pi$ to disappear. In these units, a translation by $a/4$ (as required for the diagonal translation $T$) becomes $\pi/2$, and a translation by $a/2$ (such as in $Q$) becomes $\pi$. To restore the original units, the dimensionless coordinates $x_j$ should be replaced by $2\pi x_j/a$ when necessary.

Using Eqs.~\eqref{freeBloch.eqn} and~\eqref{reciprocal.eqn} and defining
\begin{equation}
   n_1=-l_1+l_2+l_3,~~~\mbox{cyclic permutations},
\end{equation}
the Bloch phase becomes $e^{-i(n_1x+n_2 y+n_3z)}$, from which it follows that the wave functions at the $X$ point with $\vec k_X=(0,0,1)$ will be of the form
\begin{align}
   & e^{-i(n_1x+n_2 y+(n_3-1)z)},~~E_{\vec n}=\frac{\hbar^2}{2m}\epsilon_{\vec n},\label{eqnwfs}\\
   & \epsilon_{\vec n}=\left[n_1^2+n_2^2+(n_3-1)^2 \right],\label{eqnens}
\end{align}
where a dimensionless energy $\epsilon_{\vec n}$ is introduced. The conduction band of Si corresponds to the dimensionless energy $\epsilon_{\vec n}=5$, which is generated by the following 8 sets of integers:
\begin{align}
   & (200), (\bar 200), (020), (0\bar 2 0),\\
   & (202), (\bar 202), (022), (0\bar 2 2).
\end{align}
These generate the plane wave states $\psi=e^{iz}u=e^{-i(p_1x+p_2y+p_3z)}$, where the 8 degenerate states specified by the integers $(p_1,p_2,p_3)$ correspond to:
\begin{align}
   & (20\bar 1), (\bar 20\bar 1), (02\bar 1), (0\bar 2\bar 1 ),\label{firstline.eqn}\\
   & (201), (\bar 201), (021), (0\bar 21 ).\label{secondline.eqn}
\end{align}
From these integers (by summing the squares of each triple), it is clear that they all correspond to the dimensionless energy $\epsilon_{\vec n}=5$.

Let us use the above plane waves to construct two functions that transform according to the $X_1$ representation. The projection theorem~\cite{DresselhauseGroupTheory} for projecting into the (real) $X_1$ representation gives:
\begin{equation}
   \psi^{X_1}=\sum_{A\in G} \chi^{X_1}(A)~ A\psi,
\end{equation}
where $\psi$ is an arbitrary function being projected into the $X_1$ representation by applying all symmetry operations $A$ in the group $G$, weighted by the real characters $\chi(A)$ taken from the $X_1$ representation. The non-zero characters $\chi$ of the $X_1$ representation in Table~\ref{XIRtab} yield the combination ${\cal C}_1+{\cal C}_3+{\cal C}_9-{\cal C}_{14}-{\cal C}_{13}-{\cal C}_{11}$, where an overall factor of $2$ (irrelevant for symmetry properties) has been dropped. The effect of these class of operators on an arbitrary function $\psi$ simplifies as follows:
\begin{eqnarray}
   \psi^{X_1}&=&\left[ {\cal C}_1+{\cal C}_3+{\cal C}_9-{\cal C}_{14}-{\cal C}_{13}-{\cal C}_{11}\right]\psi,\nn\\
	     &=&\left[ {\cal C}_1+{\cal C}_3+{\cal C}_9-Q{\cal C}_{1}-Q{\cal C}_{2}-Q{\cal C}_{9}\right]\psi,\nn\\
	     &=&\Phi-Q\Phi,\label{X1proj.eqn}\\
	     \Phi&=&\left[E+C_{4z}^2+m_1+m_2 \right]\psi\label{semiproj.eqn}. 
\end{eqnarray}
We note that any $\psi$ selected from the set of integers in Eq.~\eqref{firstline.eqn} is proportional to $e^{-iz}$. Since the set of operations with non-zero character in Eq.~\eqref{semiproj.eqn} do not change the sign of $z$, the resulting $\Phi$ will also remain proportional to $e^{-iz}$. Likewise, any function chosen from~\eqref{secondline.eqn} will yield a $\Phi$ whose $z$-dependence is $e^{+iz}$. These two sets are related by $z\to -z$ and thus faithfully represent the two valleys. To examine the $x,y$ dependence, we select the plane wave $(p_1,p_2,p_3)=(2,0,1)$ from Eq.~\eqref{secondline.eqn}, corresponding to $\psi=e^{-i(2x+z)}$. Similarly, $(0,2,\bar1)$ corresponds to $e^{-i(2y-z)}$. The computations are summarized in Table~\ref{tabprojection}.
\begin{table}[bt]
   \centering
   \begin{tabular}{|c|c|c|}
       \hline
       $(p_1p_2p_3)$	& $(201)$ 		& $(02\bar 1)$\\ 
       \hline
       $\psi$ 		&$e^{-i(2x+z)}$		&$e^{i(z-2y)}$  \\
       \hline
       $E\psi$       	&$e^{-i(2x+z)}$    	&$e^{i(z-2y)}$\\
       $C_{4z}^2\psi$  	&$e^{-i(-2x+z)}$   	& $e^{i(z+2y)}$\\
       $m_1\psi$	&$e^{-i(+2y+z)}$	& $e^{i(z-2x)}$\\
       $m_2\psi$	&$e^{-i(-2y+z)}$	& $e^{i(z+2x)}$\\
       \hline
       $\Phi$		&$[\cos2x+\cos2y]e^{-iz}$	&$[\cos2x+\cos2y]e^{iz}$ \\
       $Q\Phi$		&$-[\cos2x+\cos2y]e^{-iz}$ 	&$-[\cos2x+\cos2y]e^{iz}$\\
       \hline
       $\psi^{X_1}$	&$[\cos2x+\cos2y]e^{-iz}$	&$[\cos2x+\cos2y]e^{iz}$\\
       \hline
   \end{tabular}
   \caption{Effect of the projection operators for the $X_1$ representation of the wave vector group at the $X$-point on representative plane waves.}
   \label{tabprojection}
\end{table}
Note that overall factors have been dropped from the expressions, as we are not normalizing the wave functions. Therefore, the following two functions generate the $X_1$ representation:
\begin{equation}
     X_1^\pm=
   (\cos2x+\cos2y) e^{\pm iz}
     \label{mybasis.eqn}
\end{equation}
Since the above set forms a two-dimensional representation of two degenerate states, the basis is determined only up to a unitary transformation. Adding and subtracting the above two functions yields the result given in Ref.~\cite{Jones1959}, namely:
\begin{equation}
  \begin{bmatrix} \tilde X^+_1\\ \tilde X^-_1  \end{bmatrix} =
  (\cos2x+\cos2y)
  \begin{bmatrix} \sin z\\ \cos z  \end{bmatrix}. 
  \label{Jonesbasis.eqn}
\end{equation}
The latter choice has the property that both components are real and the functions possess definite parity under $m_z: z\to \bar z$, but it is not compatible with the one-dimensional representations along the $\Delta$-line connecting the $X$-point to the Brillouin zone center $\Gamma$.
\begin{table*}[tb]
    \centering
    \begin{tabular}{|c|c|c|}
       \hline
       matrix & wave-vector group operator & class\\
       representation &  & \\
    \hline
    $\tau_0$ & $\{E\},\{C^2_{4z}\},\{m_1,m_2\},\{-Qm_1,Qm_2\},\{-QC^2_{4z}\},\{-Q\}$ & ${\cal C}_1,{\cal C}_3,{\cal C}_9,{\cal C}_{11},{\cal C}_{13},{\cal C}_{14}$\\
    \hline 
    $\tau_1$ & $\{C^2_{4x},C^2_{4y},-QC^2_{4x},-QC^2_{4y}\},\{S_{4x},S^{-1}_{4x},QS_{4x},QS^{-1}_{4x}\}$ & ${\cal C}_2,{\cal C}_{10}$ \\
    \hline
    $\tau_2$ & $\{gm_2,-Qgm_1\},\{-g,Qg\},\{gC^2_{4z},-QgC^2_{4z}\},\{gm_1,-Qgm_2\}$ & ${\cal C}_4,{\cal C}_6,{\cal C}_8,{\cal C}_{12}$\\
    \hline
    $\tau_3$ & $\{igS_{4z},igS^{-1}_{4z},-iQgS_{4z},-iQgS^{-1}_{4z}\},\{igC^2_{4x},igC^2_{4y},-iQgC^2_{4x},-iQgC^2_{4y}\}$ & ${\cal C}_5,{\cal C}_7$\\
    \hline
    \end{tabular}
    \caption{Representation of the operators of the wave vector group at the $X$ point of the diamond structure in the $X_1$ representation by Pauli matrices. Note that here $i$ denotes $e^{i\pi/2}$.}
    \label{valleytrans.tab}
\end{table*}

The fact that our choice of basis in Eq.~\eqref{mybasis.eqn} is the "appropriate" combination corresponding to the Si valleys can be justified not only by their inter-change through the mirror reflection $m_z$, but also by finding a combination that satisfies the compatibility relations with the $\Delta_1$ and $\Delta_{2'}$ bands. Along the $\Delta$-line, wave vectors are of the form $\vec k_\Delta=(0,0,1-\delta)$, where $\delta=1$ corresponds to the $\Gamma$ point and $\delta=0$ to the $X$ point. Similar to Eqs.~\eqref{eqnwfs} and~\eqref{eqnens}, the free electron wave functions and energies are given by:
\begin{eqnarray}
   && e^{-i(n_1x+n_2 y+(n_3-1+\delta)z)},~~E_{\vec n}=\frac{\hbar^2}{2m}\epsilon_{\vec n}\label{eqnwfs2}\\
   && \epsilon_{\vec n}=\left[n_1^2+n_2^2+(n_3-1+\delta)^2 \right]\label{eqnens2}
\end{eqnarray}
The above plane waves can be written as $\psi=e^{-i\delta z}e^{-i(p_1x+p_2y+p_3z)}$, where the triple $(p_1p_2p_3)$ are given by Eqs.~\eqref{firstline.eqn} and~\eqref{secondline.eqn}. Let us choose $(201)$ and project the resulting plane wave into the $M_1$ ($\equiv \Delta_1$) representation of the $C_{4v}$ space group of the $\Delta$-line. The starting plane wave is:
\begin{equation}
   \psi_{201}=e^{-i\delta z}e^{-i(2x+z)}. 
\end{equation}
It is evident that none of the symmetry operations of the $\Delta$-line can reverse the momentum of the state along the $z$-direction. Therefore, $e^{-iz}$ never mixes with $e^{+iz}$ under the symmetry operations of the $\Delta$-line. Consequently, it is impossible to produce $\sin z$ or $\cos z$ functions along this line. The projection into the $C_{4v}$ irreducible representations must therefore leave the $z$-momentum invariant; that is, the projected state will be proportional to $e^{-i\delta z}e^{-iz}$. Note that the non-symmorphic fractional translations can only multiply the plane waves by phase factors and cannot reverse the sign of $z$. Thus, the $\delta\to 0$ limit, which reduces to one component of the $X_1$ representation, is only compatible with the basis given in Eq.~\eqref{mybasis.eqn}.

\section{Transformation of valley degree of freedom}
\label{secBasisTransform}
Using the explicit basis functions in Eq.~\eqref{mybasis.eqn} for the $X_1$ bands that spawn the Si valleys, we can now precisely define the valley degrees of freedom $\tau_j$ in the valley space as
\begin{align}
        &\hat\tau_1=\begin{bmatrix}
        |X^+_1\rangle & | X^-_1\rangle 
    \end{bmatrix} 
    \begin{bmatrix}
        0 & 1 \\
        1 & 0
    \end{bmatrix}
    \begin{bmatrix}
        \langle X^+_1| \\ \langle X^-_1|
    \end{bmatrix},\label{t1.eqn}\\
     &\hat\tau_2=\begin{bmatrix}
         |X^+_1\rangle & |X^-_1\rangle 
    \end{bmatrix} 
    \begin{bmatrix}
        0 & -i \\
        i & 0
    \end{bmatrix}
    \begin{bmatrix}
         \langle X^+_1| \\ \langle X^-_1| 
    \end{bmatrix},\label{t2.eqn}\\
     &\hat\tau_3=\begin{bmatrix}
         |X^+_1\rangle & |X^-_1\rangle 
    \end{bmatrix} 
    \begin{bmatrix}
        1 & 0 \\
        0 & -1
    \end{bmatrix}
    \begin{bmatrix}
         \langle X^+_1| \\ \langle X^-_1| 
    \end{bmatrix}. \label{t3.eqn}
\end{align}
Let us emphasize again that we use numeric subscripts for the valley Pauli matrices to indicate that—unlike the spin components $\sigma_x,\sigma_y,\sigma_z$—they are not associated with the spatial directions $x$, $y$, or $z$.

The first observation from Eq.~\eqref{mybasis.eqn} is that under time reversal (TR) $X_1^\pm \to X_1^\mp$, meaning  the two valleys are exchanged-- an intuitively result. Applying this transformation to the valley degrees of freedom~\eqref{t1.eqn}-\eqref{t3.eqn} we find that under TR
\begin{equation}
    \hat\tau_1\to \hat\tau_1,~~~\hat\tau_2\to \hat\tau_2,~~~\hat\tau_3\to -\hat\tau_3. \label{tauTR.eqn}
\end{equation}
This behavior sharply contrasts with that of spin Pauli matrices, which all change sign under TR. For the valley Pauli matrices, only $\hat\tau_3$ reverses sign.

Having the explicit form in Eq.~\eqref{mybasis.eqn} for the functions $X^+_1$ and $X^-_1$ allows us to determine how the wave vector group at $X$ acts on the valley degrees of freedom defined in Eqs.~\eqref{t1.eqn}-\eqref{t3.eqn}.  To do this, we apply all 32 operators of the wave vector group to the coordinates $(x,y,z)$, from which the corresponding transformations of the $X^\pm_1$ functions are constructed. 
Indeed, the sharp contrast between the behavior of spin and valley degrees of freedom under TR in Eq.~\eqref{tauTR.eqn} suggests that the transformation of the $\hat \tau_i$ matrices under the wave vector group must also be entirely different from that of spins.

Starting with the basis valley states in Eq.~\eqref{mybasis.eqn}, and using units where $a=2\pi$, the translation $T=a(1,1,1)/4$ becomes $\frac{\pi}{2}(1,1,1)$. Since the transverse coordinates in Eq.~\eqref{mybasis.eqn} appear only as $2x$ and $2y$, the translation $T$ can be represented as:
 \begin{equation}
     T:\begin{bmatrix}
         2x \\ 2y \\z
     \end{bmatrix}\to\begin{bmatrix}
         2x+\pi \\ 2y+\pi \\ z+\frac{\pi}{2}
     \end{bmatrix}.
     \label{translation.eqn}
 \end{equation}
Similarly, the face-diagonal translation operator $Q=a(0,\bar 1,\bar 1)/2$ becomes a translation by $\pi(0,\bar 1,\bar 1)$, which induces:
 \begin{equation}
     Q   :\begin{bmatrix}
         2x \\ 2y \\z
     \end{bmatrix}\to\begin{bmatrix}
         2x+0 \\ 2y-2\pi \\ z-\pi
     \end{bmatrix}\equiv
     \begin{bmatrix}
         2x\\ 2y \\ z-\pi 
     \end{bmatrix}.
     \label{Qop.eqn}
 \end{equation}
One must also include the mirror reflection in the $xy$ plane, $m_z:xyz\to xy\bar z$, which completes the construction of $g=Tm_z$. The effect of other point group elements involved in Eq.~\eqref{C114.eqn} is explicitly given by Eq.~\eqref{pointgroup.eqn}.

Using the above rules, we can construct the transformation of the basis functions in Eq.~\eqref{mybasis.eqn}. The results are summarized in Table~\ref{valleytrans.tab}, where $i=e^{i\pi/2}$. Let us work through a few entries of this table. For example, for the $S_{4z}$ operator belonging to class ${\cal C}_{10}$, the coordinates $(x,y,z)$ are transformed to $(\bar y, x,\bar z)$. This transformation leaves the $\cos2x+\cos2y$ part unchanged but converts $e^{\pm iz}$ to $e^{\mp iz}$. Therefore, the $S{4z}$ operation flips the valley,
\begin{align}
   S_{4z} X_1^\pm= X_1^\mp \rightarrow S_{4z} \dot{=} \tau_1,
\end{align}
where the symbol $\dot =$ means "is represented by". The operator $S^{-1}_{4z}$ is similarly represented by the Pauli matrix $\tau_1$. For the operator $QS_{4z}$, the additional operation $Q$ introduces an extra shift in $z$ according to Eq.~\eqref{Qop.eqn}, which changes the $z$ dependence as $e^{\pm iz}\to e^{\mp i(z-\pi)}=-e^{\mp iz}$. Therefore:
\begin{align}
   QS_{4z} X_1^\pm = - X_1^\mp \rightarrow QS_{4z} \dot = -\tau_1. 
\end{align}
As another example, consider the operators of class ${\cal C}_{12}$. For $Qgm_2$, the coordinates $(x,y,z)$ are transformed to $Q(\bar y+\pi/2,\bar x+\pi/2,\bar z+\pi/2)$, or equivalently $(2x,2y,z)\to (2\bar y+\pi,2\bar x+\pi,\bar z-\pi/2)$. The change in sign of $z$ implies that the operator will be represented by an off-diagonal Pauli matrix. This transformation indeed has the following effect:
\begin{align}
   Qgm_2 X_1^\pm = e^{\pm i\pi/2}X_1^\mp = \pm i X_1^\mp\rightarrow Qgm_2\dot= -\tau_2.
\end{align}
A similar computation gives $gm_1\dot = \tau_2$. In this way, the explicit representation of all 32 members of the wave vector group at the $X$ point is constructed and summarized in Table~\ref{valleytrans.tab}.
\begin{table*}[btp]
    \centering
    \begin{tabular}{|c|c|c|c|c|c|c|c|c|c|c|c|c|c|c|l|}
       \hline 
             & $M_1$ & $M_2$ & $M_3$ & $M_4$ & $M_5$ & $M'_1$ & $M'_2$ & $M'_3$ & $M'_4$ & $M'_5$ & $X_1$ & $X_2$ & $X_3$ & $X_4$ & Objects\\
       \hline
       $M_1$ & $M_1$ & & & & & & & & & & & & & &$\tau_0,k_z^2,k_x^2+k_y^2,\varepsilon_{zz}$  \\
       \hline
       $M_2$ & $M_2$ & $M_1$ & & & & & & & & & & & & &$k_x^2-k_y^2,\varepsilon_{xx}-\varepsilon_{yy}$  \\
       \hline
       $M_3$ & $M_3$ &$M_4$ &$M_1$ & & & & & & & & & & & & $R_z$ \\
       \hline
       $M_4$ & $M_4$ &$M_3$ &$M_2$ &$M_1$ & & & & & & & & & & & $\tau_2,k_xk_y,\varepsilon_{xy}$\\
       \hline
       $M_5$ & $M_5$ &$M_5$ &$M_5$ &$M_5$ &$M_1+M_2$ & & & & & & & & & &$(k_x,k_y),(P_x,P_y)$\\
        & & & & &$M_3+M_4$ & & & & & & & & & &$(\partial_x,\partial_y)$\\
       \hline
       $M_1'$ & $M'_1$ &$M'_2$ &$M'_3$ &$M'_4$ &$M'_5$ &$M_1$ & & & & & & & & &$P_xR_x+P_yR_y$\\
       \hline
       $M_2'$ & $M'_2$ &$M'_1$ &$M'_4$ &$M'_3$ &$M'_5$ &$M_2$ &$M_1$ & & & & & & & &$\tau_1,P_xR_x-P_yR_y$\\
       \hline
       $M_3'$ & $M'_3$ &$M'_4$ &$M'_1$ &$M'_2$ &$M'_5$ &$M_3$ &$M_4$ &$M_1$ & & & & & & & $\tau_3,k_z,\partial_z,P_xR_y-P_yR_x$\\
       \hline
       $M_4'$ & $M'_4$ &$M'_3$ &$M'_2$ &$M'_1$ &$M'_5$ &$M_4$ &$M_3$ &$M_2$ &$M_1$ & & & & & &$P_xR_y+P_yR_x$\\
       \hline
       $M_5'$ & $M'_5$ &$M'_5$ &$M'_5$ &$M'_5$ &$M'_1+M'_2$ &$M_5$ &$M_5$ &$M_5$ &$M_5$ &$M_1+M_2$ & & & & &$(k_xk_z,k_yk_z),(R_x,R_y)$\\
	&  & & & &$M'_3+M'_4$ & & & & & $M_3+M_4$ & & & & &$(\varepsilon_{xz},\varepsilon_{yz})$\\
       \hline
       $X_1$ & $X_1$ &$X_2$ &$X_2$ &$X_1$ &$X_3+X_4$ &$X_2$ &$X_1$ &$X_1$ &$X_2$ &$X_3+X_4$ &$M_1+M_4$ & & & & $(X_1^+,X_1^-)$\\
        &  & & & & & & & & & &$M'_2+M'_3$ & & & &\\
       \hline
       $X_2$ & $X_2$ &$X_1$ &$X_1$ &$X_2$ &$X_3+X_4$ &$X_1$ &$X_2$ &$X_2$ &$X_1$ &$X_3+X_4$ &$M_2+M_3$ &$M_1+M_4$ & & &\\
        &  & & & & & & & & & &$M'_1+M'_4$ &$M'_2+M'_3$ & & &\\
      \hline
       $X_3$ & $X_3$ &$X_4$ &$X_4$ &$X_3$ &$X_1+X_2$ &$X_3$ &$X_4$ &$X_4$ &$X_3$ &$X_1+X_2$ &$M'_5+M_5$ &$M'_5+M_5$ &$M_1+M_4$ & &\\
        &  & & & & & & & & & & & &$M'_1+M'_4$ & &\\
       \hline
       $X_4$ & $X_4$ &$X_3$ &$X_3$ &$X_4$ &$X_1+X_2$ &$X_4$ &$X_3$ &$X_3$ &$X_4$ &$X_1+X_2$ &$M'_5+M_5$ &$M'_5+M_5$ &$M_2+M_3$ &$M_1+M_4$ &\\
        & & & & & & & & & & & & &$M'_2+M'_3$ &$M'_1+M'_4$ &\\
       \hline
    \end{tabular}
    \caption{Multiplication table for the group of wave vector at $X$ point of the diamond lattice structure. The momentum components $k_j$ in a $k.p$
    description always are coupled to $p_j$, so they transform like derivative $\partial_j$ and hence translation parts of non-symmorphic operators 
    does not matter. $P_j$ in this table represents a typical polar vector. $R_j$ is usual the axial vector components. }
    \label{repmultiplication.tab}
\end{table*}

These explicit matrix representations allow us to emphasize again the distinction between the valleyor basis in Eq.~\eqref{mybasis.eqn} and spinors. Consider, for example, the operator $C_{4z}^2$, a rotation by $\pi$ around the $z$-axis that belongs to class ${\cal C}_3$. Its effect on the valleyor basis is given by the identity matrix $\tau_0$, whereas for spinors, a rotation by $\pi$ around the $z$-axis is represented by the $SU(2)$ matrix $e^{i\sigma_z\pi/2}=i\sigma_z$. Hence, in the spinor representation, $C^2_{4z}$ corresponds to the third Pauli matrix, while in the valleyor representation it corresponds to the identity matrix. Therefore, it is misleading to refer to the valley degree of freedom as a pseudospin. The representation of valleys as valleyors raises interesting questions regarding the path integral quantization of valley dynamics.

Now, equipped with the explicit representation of every operator $a$ in the wave vector group by a corresponding $2\times 2$ matrix $\tau_A$, as detailed in Table~\ref{valleytrans.tab}, we can compute their characters. The transformation of, for example, $\tau_1$ will be:
\begin{align}
   \tau_1 \to \tau_A \tau_1 \tau_A^\dagger=\left \{
      \begin{array}{cc}
	 (+1) \tau_1 & \tau_A=\tau_0\mbox{ or }\tau_1\\
	 (-1) \tau_1 & \tau_A=\tau_2\mbox{ or }\tau_3\\
      \end{array}
      \right..
\end{align}
Therefore, the character of $\tau_1$ is $+1$ for classes $1,2,3,9,10,11,13,14$ and $-1$ otherwise. This matches the $M_2'$ representation, i.e., $\tau_1 \in M_2'$, or equivalently, the symmetry type of $\tau_1$ is $M'_2$. Similarly, from Table~\ref{valleytrans.tab}, the symmetry type of the other valley components can be identified. This leads to:
\begin{align}
   \tau_1 \in M_2',~~~\tau_2\in M_4,~~~\tau_3\in M_3'
   \label{main.eqn}
\end{align}
It is interesting to note that $\tau_2$ is an $M$-type object (even under $m_z$ reflection), whereas $\tau_1$ and $\tau_3$ are $M'$-type, i.e., they are odd under $m_z$ reflection. This behavior is markedly different from that of spinors, for which the $z(y)$-component $\sigma_{z(y)}$ of spin is even (odd) under $m_z$ reflection.

Equation~\eqref{main.eqn}, derived from the transformation properties of the valleyor basis, is incorporated into Table~\ref{valleytrans.tab}. This equation is the central result of our paper, enabling us to determine how the valley Pauli matrices couple to various vector and tensor fields. The only remaining task is to assign a symmetry type (irreducible representation) to the components of the desired vector/tensor field and to construct the multiplication table for the irreducible representations of the wave vector group at the $X$ point of the silicon structure, which is done in the next section.

\section{Symmetry-allowed terms in the effective Hamiltonian: Topological Nodal Line}
\label{secCoupling}
In Table~\ref{repmultiplication.tab}, we have computed the multiplication table for all 14 representations by projecting the character product into the irreducible representations. The last column is constructed for a generic macroscopic polar vector $P_j=(P_x,P_y,P_z)$ that is insensitive to the translation part of the space group. As such, it can represent the derivative of any quantity that is scalar with respect to the group of the $X$-point. Examples include the electric field, which is $-\partial_j\phi$ where $\phi$ is electrostatic potential.  The momentum $k_j$ is another legitimate example of $P_j$, since in a $\bf{k} \cdot \bf{p}$ framework, $k_j$ always accompanies a derivative $-i\hbar \partial_j$, making fractional translations of the space group irrelevant. In quantum dot geometries, $P_j$ could also represent the electric polarization, which indicates a preferred direction and can be produced by \emph{gate asymmetry.} In shuttling geometries~\cite{Kuenne2024}, asymmetry between the left and right sides of the shuttling path can be a source of a $P$ with a component transverse to the path.

The $R_j$s are axial vector components, i.e., objects of the form $\epsilon_{jkl}P_kV_l$, where both $P$ and $V$ are polar vectors. They represent the antisymmetric part of the tensor $P_kV_l$. The most significant examples relevant to our problem are the magnetic field $B_j$ or magnetization. In quantum dots, an additional axial vector can arise from the dot's ellipticity, specified by an axial vector $n_j$ along the elongation axis.

Equipped with Table~\ref{repmultiplication.tab}, we can now construct invariant terms. Let us begin by examining the possible terms in the band structure itself. This approach leads to the identification of a significant new term that has been overlooked in the semiconductor community.

\subsection*{Topological nodal line k$\cdot$p term}
A powerful application of the method of invariants~\cite{Willatzen2009} is to determine permissible terms describing the band structure itself. Using Table~\ref{repmultiplication.tab}, we can write down the band structure of pure silicon. One can immediately see that the well-known shifted parabolas of the silicon valleys correspond to $M_1\times M_1$ and $M'_3\times M'_3$ invariants from the $M_1$ and $M'_3$ channels. The $M_1\times M_1$ invariant contributes two identical parabolas, while the $M'_3\times M'_3$ invariant introduces a right/left shift corresponding to the two eigenvalues of the $\tau_3$ matrix from the $M'_3$ representation, as follows:
\begin{align}
   \left(\frac{k_z^2}{2m_\ell}+\frac{k_x^2+k_y^2}{2m_t} \right)\tau_0+\frac{\hbar k_1}{m_\ell}\tau_3 k_z,
   \label{basicparabolas.eqn}
\end{align}
where we have identified the coupling constants of the $k_z^2$ and $k_x^2+k_y^2$ terms from the $M_1$ representation with the inverse longitudinal and transverse effective masses, and the coupling constant of the $k$-linear term ($M'_3$ representation) is identified with the displacement $\pm k_1$ of the valleys relative to the $X$ point. However, this picture of two parabolas shifted by $\pm k_1$ from the $X$ point is not the complete story.

It is immediately apparent that at the quadratic level (second power of momenta $k$), the $M_4$ channel would also contribute a $k_xk_y$ term of $M_4$ symmetry type. One therefore expects an invariant of the $M_4\times M_4$ type that naturally selects the matrix $\tau_2$ in the valley space to form:
\begin{align}
   \kappa k_x k_y\tau_2, 
   \label{nodalline}
\end{align}
where the coupling constant $\kappa$ has dimensions of inverse effective mass. This term, being quadratic in momentum, is equally important as the other quadratic terms in Eq.~\eqref{basicparabolas.eqn}. It indicates that the $\Delta_1$ and $\Delta_{2'}$ bands forming the silicon valleys cross along the entire line connecting the $X$ point to the $W$ points (see the cyan and black points in Fig.~\ref{bands.fig}). The $XW$ line, also known as the $Z$ line, is known have two-dimensional irreducible representation that preserves the two-fold degeneracy of the $X$-point~\cite{DresselhauseGroupTheory,Koster1957}. Our analysis has the advantage of revealing the precise form—namely, the $k_xk_y$ dependence—according to which the separation between the two bands vanishes along the $XW$ line ($k_x=0$ or $k_y=0$). Indeed, this quadratic form can be confirmed by projecting the tight-binding bands of silicon onto the two-fold degenerate manifold of $X_1$ representation~\cite{Chamon2020}, and the value of $\kappa$ (in units where $\hbar$ and $a$ are set to 1) is $0.18$ eV. In the same units, $(2m_l)^{-1}$ is $0.132$ eV; thus, the $\kappa$ term, being $1.363$ times stronger than $(2m_l)^{-1}$, cannot be neglected within the family of quadratic terms. It also has topological significance, as the network of $Z$ lines forms a semi-Dirac nodal line structure that contributes a Berry flux of $\pi$~\cite{Chamon2020}. Therefore, the total effective Hamiltonian for pure Si becomes:
\begin{align}
   \left(\frac{k_z^2}{2m_\ell}+\frac{k_x^2+k_y^2}{2m_t} \right)\tau_0+\frac{\hbar k_1}{m_\ell}\tau_3 k_z+\kappa k_x k_y \tau_2,
   \label{ourmodel.eqn}
\end{align}
Note that we have obtained this term through pure theoretical reasoning centered on the $X$ point. In doing so, we not only recover the vanishing of the $\Delta_1$-$\Delta_{2'}$ separation along the network of $XW$ lines, but also determine the precise $k_xk_y$ dependence of this vanishing, which is consistent with integrating out other bands to second order in $k$ to compute the coupling $\kappa$ in a tight-binding approach~\cite{Chamon2020}. The nodal line structure is consistent with the existence of two-dimensional representations along the $Z$ line~\cite{DresselhauseGroupTheory,Koster1957}, which gives rise to a $\pi$ Berry flux~\cite{Chamon2020}. Therefore, the non-symmorphic structure of the wave vector group at the $X$-point not only determines how the silicon valley states are spawned by the valleyor basis $X_1^\pm$, but also imposes the nodal-line structure along the $XW$ lines, giving the precise d-wave form $k_xk_y$ and fixing the corresponding valley Pauli matrix to be $\tau_2$. This insight has significant implications for constructing effective microscopic models of the conduction band in silicon.

To compare our derivation with Ref.~\cite{Chamon2020}, the growth direction must first be set to the $x$-direction. This \emph{does not} change the Pauli matrix $\tau_3$ to $\tau_1$, as the valley Pauli matrices (unlike spin Pauli matrices) are not tied to the $x,y,z$ directions. To compare our Hamiltonian with that in Ref.~\cite{Chamon2020}, one simply needs to introduce the combinations $|a\rangle=(|+\rangle+|-\rangle)/\sqrt{2}$ and $|b\rangle=(|+\rangle-|-\rangle)/\sqrt{2}$ for the valleyor basis states $|\pm\rangle$. This change of valley basis maps $\tau_3$ to $\tau_1$ and vice versa, while $\tau_2$ only changes sign. In this way, our form in Eq.~\eqref{ourmodel.eqn} can be mapped to the one derived in Ref.~\cite{Chamon2020}. Thus, the basis used in Ref.~\cite{Chamon2020} are~\eqref{Jonesbasis.eqn}, and not the valley-basis~\eqref{mybasis.eqn}.

Apart from its topological significance, can this nodal-line structure serve as a new resource for coupling the valleys? Our preliminary results indicate that in quantum dot settings, a pure in-plane magnetic field can leverage the nodal line coupling $\kappa$ to induce valley splitting between two low-lying states with opposite $\langle \tau_2\rangle=\pm 1$ polarizations. The key is that the confining potential of the dot introduces a frequency scale $\omega_D$ which, in combination with the cyclotron frequency $\omega_C$, can form an energy scale comparable to the 80 meV~\cite{Chamon2020} separation between the silicon conduction band edge and the semi-Dirac point at $X$. In this way, off-diagonal matrix elements of the $\tau_2$ type can induce valley splitting. Thus, the nodal line structure can serve as an additional resource, providing a further mechanism for valley splitting.

\section{Coupling of strain to valley degrees of freedom}
\subsection{Pure strain terms}
Strain, being a tensor, allows its components to be associated with various irreducible representations in Table~\ref{repmultiplication.tab}. The strain components are defined by $\varepsilon_{ij}=(\partial_i u_j+\partial_j u_i)/2$, where $u_j$ is the $j$th component of the displacement field. As an example, the strain components $\varepsilon_{zz}$ and $\varepsilon_{xx}+\varepsilon_{yy}$ belong to the $M_1$ representation; namely, they are scalars~\footnote{Note that the behavior of $\varepsilon_{xx}+\varepsilon_{yy}$ is similar to $k_x^2+k_y^2$. }. Therefore, they can appear in the following $M_1\times M_1$ invariant forms:
\begin{align}
   \tau_0 \varepsilon_{zz},~~\tau_0\left( \varepsilon_{xx}+\varepsilon_{yy}\right).
\end{align}
These perturbations merely shift the energy of the semi-Dirac crossing\footnote{i.e., linear in the $z$ direction and quadratic in the $xy$ plane.} at the $X$ point without gapping it out. Strain can also couple in the $M_4$ channel in the form of a $M_4\times M_4$ scalar:
\begin{align}
   \tau_2 \varepsilon_{xy}.\label{shearstrain}
\end{align}
This form of shear strain gaps out the nodal degeneracy at the $X$ point. Similar strain-induced gap terms in different bases are obtained in Refs.~\cite{Hensel1965} and~\cite{Sverdlov2008}. The significance of gap terms in the modern theory of Berry phase is that they produce a non-zero orbital magnetization contribution~\cite{NiuRMP2010}. In fact, the orbital magnetization arising from non-trivial Berry structure acts like a $B$-field in $k$-space and is therefore expected to modify how an external $B$-field affects Zeeman-split spin states. Thus, the above term is expected to generate a non-trivial interplay between $g$-factors and the valleys.

The equation has a Zeeman-like form $\vec{\cal B}\cdot \vec \tau={\cal B}_2\tau_2$, where a valley-magnetic field component ${\cal B}_2=\varepsilon_{xy}$ emerges from symmetry. This field couples to the valley Pauli matrix $\tau_2$ in a manner analogous to the Zeeman interaction. It demonstrates that while the valley eigenstates of $\vec\tau$ do not behave like spinor eigenstates of $\vec\sigma$ and thus cannot couple directly to the magnetic field $\vec B$, they can utilize other background fields to generate their own valley-magnetic components ${\cal B}_j$. These components couple to $\tau_j$ just as $\vec B$ couples to $\vec\sigma$. The next sub-section will show how combinations of physical fields can produce the remaining components of the valley-magnetic field.

\subsection{Valley-magnetic fields from physical magnetic fields}

The magnetic field has three components $B_i$, with $i=x,y,z$. All components of $\vec B$ are TR-odd. However, according to Eq.~\eqref{tauTR.eqn}, only $\tau_3$ is TR-odd. Therefore, terms linear in $B$ cannot couple to $\tau_1$ or $\tau_2$. Therefore, terms such as
\begin{eqnarray}
   &&(B_x\varepsilon_{xz} + B_y\varepsilon_{yz}) \tau_0. \nonumber \\
   &&(B_x\varepsilon_{yz} + B_y\varepsilon_{xz}) \tau_2.\nonumber
\end{eqnarray}
which arise from the decomposition $M'_5\times M'_5=M_1+M_2+M_3+M_4$ in Tab.~\ref{repmultiplication.tab}, are excluded. Although they are invariant under the space group of the wave vector at the $X$ point, they are not invariant under time reversal.

Having ruled out couplings of $\tau_1$ and $\tau_2$ with odd powers of the magnetic field, our aim is now to find terms of the form ${\cal B}_3\tau_3$. Here, ${\cal B}_3$ can be interpreted as the third component of a "valley-magnetic" field—a composite field constructed from other physical fields that can afford a Zeeman-like coupling to $\tau_3$. 

To classify all such terms that include odd powers of the physical magnetic field $B_i$, we begin by noting that $\tau_3$ has $M'_3$ symmetry type. Therefore we need a combination of $B_i$ with other fields that would also behave as $M'_3$. The mathematical possibilities according to Tab.~\ref{repmultiplication.tab} are $M'_1\times M_3,M'_3\times M_1,M'_2\times M_4,M'_4\times M_2,M'_5\times M_5$. Out of these terms, the following give non-trivial candidates for ${\cal B}_3$:
\begin{align}
& M'_3\times M_1 \to (P_xB_y-P_yB_x)~\{1,\varepsilon_{zz},(\varepsilon_{xx}+\varepsilon_{yy})\},\label{tensileB3.eqn}\\
& M'_2\times M_4 \to (P_xB_x-P_yB_y)~\varepsilon_{xy},\label{shearxyB3.eqn}\\
& M'_4\times M_2 \to (P_xB_y+P_yB_x)~ (\varepsilon_{xx}-\varepsilon_{yy}).\label{shearxxyyB3.eqn}
\end{align}
For quantum dots in SiGe/Si/SiGe heterostructures, where the polar vector $\vec P$ can represent a preferred direction arising from the quantum dot's asymmetry, the meaning of the above equations is as follows: According to Eq.~\eqref{tensileB3.eqn}, when the in-plane magnetic field is perpendicular to $\vec P$, the $z$-component of $\vec P \times \vec B$ becomes a permissible candidate for ${\cal B}_3$. Tensile strain terms $\varepsilon_{zz}$ and $(\varepsilon_{xx}+\varepsilon_{yy})$ can enhance this coupling. 
On the other hand, according to Eqs.~\eqref{shearxyB3.eqn} and~\eqref{shearxxyyB3.eqn}, the shear strain components $\varepsilon_{xy}$ and $(\varepsilon_{xx}-\varepsilon_{yy})$ can be leveraged when the in-plane magnetic field $\vec B$ is aligned with the polar vector $\vec P$.
In other words, when substantial shear strain is present, the magnetic field should be aligned along the quantum dot's asymmetry vector $\vec P$. Conversely, when substantial compressive strain is present, the in-plane magnetic field must be aligned perpendicular to the asymmetry vector $\vec P$ of the quantum dot to maximize the benefit from the background strain fields. 
In a forthcoming publication, we show that the optimal valley splitting from the topological nodal line term in Eq.~\eqref{nodalline} occurs when the in-plane magnetic field is directed at $45^\circ$ to the crystallographic axes. Therefore, the combined message of the above equations is that the quantum dot's asymmetry vector $\vec P$ should always be oriented at $45^\circ$ relative to the $x$ and $y$ crystallographic axes. Depending on the nature of the background strain, the in-plane magnetic field can then be chosen to be either parallel or perpendicular to the asymmetry direction $\vec P$.

There is yet another possible coupling in tilted magnetic fields that, although nonlinear in the $B$-field, is worth investigating, as a $B_z$ component can arise from field misalignment in realistic experimental situations. The question is whether such a misalignment is beneficial or detrimental. If a magnetic field possesses both in-plane and out-of-plane components, its $z$-component $B_z$ belongs to the $M_3$ representation and can be combined with other background fields. All representation products capable of coupling to the $M_3$ symmetry type to form a scalar are $M'_1 \times M'_3$, $M_2 \times M_4$,  $M_5 \times M_5$, $M'_5 \times M'_5$ and $M'_4\times M'_2$. The resulting terms from these coupling channels are as follows:
\begin{itemize}[label=$\circ$]
    \item The combination $(P_xB_x+P_yB_y)$, belonging to the $M'_1$ representation, and $\tau_3\in M'_3$, produces the term $B_z(P_xB_x+P_yB_y)\tau_3$. Since this term is TR-odd, it must be excluded.
    \item $M_2 \times M_4\to B_z(\varepsilon_{xx}-\varepsilon_{yy})\tau_2$ is again TR-odd and must be excluded.
    \item $M_5\times M_5$ and $M'_5\times M'_5$ contain no $\tau$ degree of freedom and must be excluded.
\end{itemize}
Therefore, the only remaining TR-invariant possibility is the combination
\begin{equation}
M_3\times M'_4\times M'_2\to B_z(P_xB_y+P_yB_x)\tau_1,
\end{equation}
which corresponds to the valley-magnetic field component
\begin{equation}
{\cal B}_1 \propto B_z(P_xB_y+P_yB_x).
\label{nlB1}
\end{equation}
This equation indicates that a slight out-of-plane tilt of the magnetic field need not be avoided. When the in-plane component of the magnetic field is perpendicular to the in-plane asymmetry vector $\vec P$, this tilt can generate the first component ${\cal B}_1$ of the valley-magnetic field.

The possible valley-magnetic fields that can be obtained from the physical fields $\vec P,\vec B,\varepsilon_{ij}$ are summarized in Tab.~\ref{VMF.tab}.
\begin{table}[hbt]
\begin{tabular}{|c|l|}
    \hline 
    ${\cal B}_1$ & $B_z(P_xB_y+P_yB_x)$ \\
    \hline 
    ${\cal B}_2$ & $\varepsilon_{xy}$  \\
    \hline 
    ${\cal B}_3$ & $(P_xB_y-P_yB_x)\{1,\varepsilon_{zz},(\varepsilon_{xx}+\varepsilon_{yy})\}$\\
    & $(P_xB_x-P_yB_y)\varepsilon_{xy},~ (P_xB_y+P_yB_x)(\varepsilon_{xx}-\varepsilon_{yy})$\\
    \hline 
\end{tabular}
\caption{Valley-magnetic field components ${\cal B}_i$ in terms of physical background fields.}
\label{VMF.tab}
\end{table}

\subsection{Fluctuations} 
In Ref.~\cite{paquelet_wuetz_atomic_2022}, it has been demonstrated that atomic fluctuations in the Ge concentration produce non-deterministic valley splitting. The underlying mechanism is that scattering from the random scalar potential created by Ge atoms generates off-diagonal matrix elements between the two valley states~\cite{Friesen2007}. Within this picture: (1) the effect is proportional to $\tau_1$, which, combined with the $U(1)$ phase arising from the Bloch phase of different layers, leads to a Rice distribution of valley splittings; (2) the magnitude of this effect is fixed for a given device and is determined by the specific profile of the Ge atoms.

The three types of valley-magnetic fields derived in this work, summarized in Tab.~\ref{VMF.tab}, indicate two important implications. First, a background ${\cal B}_3$ can be produced that directly couples to $\tau_3$. This extends the parameter space for valley splitting from entities arising from \emph{scalar} potential scattering—which couples only to $(\tau_1,\tau_2)$—to those arising from \emph{vector} potentials, which in a three-dimensional space can couple to $(\tau_1,\tau_2,\tau_3)$. The third component ${\cal B}_3$, as a new resource capable of coupling to $\tau_3$, is promising for the following reason: in regions where the first two components coupling to $(\tau_1,\tau_2)$ produce a vanishingly small valley splitting—responsible for the left tail of the Rice distribution—the ${\cal B}_3$ component can prevent the valley splitting from vanishing. This effectively shifts the tail of the Rice distribution away from zero, thereby avoiding instances of negligible valley splitting.

Second, assuming a fixed background in-plane magnetic field $(B_x,B_y)$, the vector $(P_x,P_y)$ can represent a fluctuating electric field arising from random Ge atoms. The combination of the background $\vec B$ and fluctuating $\vec P$ can leverage both tensile and shear strain. This strain can originate from both a background (average) value due to external stressors and a fluctuating component from random Ge atoms. The random component is always present for randomly distributed Ge atoms, while a non-zero average strain can be externally applied. Note that according to Eq.~\eqref{tensileB3.eqn}, even without any externally applied strain, a ${\cal B}_3$ field can still be generated by the combination of $\vec B$ and the (random) $\vec P$ field. The effect of tensile strain is to enhance the magnitude of this resulting ${\cal B}_3$ field. 

The third characteristic of the valley-magnetic mechanism for valley splitting is its linear dependence on each of the fields: $\vec B$, $\vec P$, and $\varepsilon_{ij}$. This implies that doubling either the applied $\vec B$ or the externally applied strain $\varepsilon_{ij}$ will also double the contribution of the valley-magnetic field channel to the valley splitting. This property can serve as a means to verify the existence of such terms and to separate their contribution from that of the scalar potential scattering terms to the total valley splitting.


\section{Additional resources in quantum dots}

How can this construct be helpful in quantum dots that confine spin qubits? We begin by noting that spin, while fundamentally a property of the full rotation group, remains a useful attribute of electrons even in confined geometries with significantly reduced symmetries. This is precisely why a spin isolated by quantum dot confining potential can be utilized for quantum computation. Similarly, although the valley degree of freedom—as shown in the present work—differs fundamentally from spin or pseudospin, it is expected to remain a meaningful and useful degree of freedom in confined geometries. Therefore, just as fundamental couplings like the Zeeman interaction for spins persist in confined geometries (albeit with variations in the $g$-factor), our valley-magnetic coupling, arising from the sources summarized in Table~\ref{VMF.tab}, is also expected to survive in quantum dots. This may introduce a valley-magnetic renormalization factor $g_{\rm vm}$, analogous to the $g$-factors for spins. Since reducing symmetries generally imposes fewer constraints, additional couplings—such as those induced by dot ellipticity, discussed below—may also become possible.

The shape and size of a quantum dot inherently defines its properties and plays a crucial role by breaking specific symmetries. In addition to random sources of $\vec P$, the shape asymmetry of the quantum dot can itself be described by a polar vector $(P_x,P_y)$. This provides a deterministic source of polar vector components that can be leveraged—in conjunction with a magnetic field or its combination with tensile or shear strain—as detailed in table~\ref{VMF.tab}.
Furthermore, in a shuttling setup where an electron is transported along the $x$ direction, a lateral asymmetry can induce a persistent $P_y$ component along the entire shuttling path. This asymmetry can activate all valley-magnetic couplings in table~\ref{VMF.tab} that involve $P_y$. To test for the presence of these terms, one could compare the statistics of valley splitting in setups like that of Ref.~\cite{volmer_mapping_2024} for shuttling paths located at the middle of the channel versus paths near its edges.

Another aspect of the quantum dot's shape (that cannot be described by a polar vector) is its ellipticity. This property is specified by a major axis, where both directions along this axis are equivalent. This symmetry can be encoded into a nematic-like order parameter, which in two dimensions is given by the traceless tensor~\cite{chaikin2000}
\begin{equation}
Q_{ab}=n_an_b-\frac{1}{2}\delta_{ab},
\end{equation}
where $a,b=x,y$. Note that this quantity does not arise from electron-electron interactions~\cite{Nie2014} but is instead externally imposed by the shape of the quantum dot, insofar as that shape can be modeled by ellipticity. This quantity is invariant under $n_a \to -n_a$, meaning it is even with respect to both directions along the vector $\vec n$. 
The traceless nature of this tensor ensures it has no component belonging to the $M_1$ irreducible representation, meaning it excludes any rotationally invariant part of the shape as it should. By construction, its off-diagonal part is symmetric under $x \leftrightarrow y$, leaving only two independent components: $Q_{xy}$ and $Q_{xx}-Q_{yy}$. Given their symmetry character, these behave like shear strain components. As a result, they can give rise to the following possible valley-magnetic field components:
\begin{eqnarray}
    &&{\cal B}_2:~Q_{xy},\label{Qshear.eqn}\\
    &&{\cal B}_3:~ (P_xB_x-P_yB_y)~Q_{xy},\label{QshearEB.eqn}\\
    &&{\cal B}_3:~(P_xB_y+P_yB_x)~ (Q_{xx}-Q_{yy})\label{QshearPlus.eqn}
\end{eqnarray}
These couplings represent additional mechanisms beyond possible valley-magnetic fields arising purely from background strain. The channels described above can strengthen the overall effect of strain. 
These new terms suggest a simple strategy to leverage the dot's ellipticity: Eq.~\eqref{Qshear.eqn} implies that a built-in ${\cal B}_2$ component can be constructed simply by aligning the major axis of the elliptic dot at $45^\circ$ to the crystallographic axes. In this case the tensor $Q$ becomes
\begin{equation}
    Q=\begin{bmatrix}
        0 & 0.5\\
        0.5 & 0
    \end{bmatrix}. \label{Qtensorxy.eqn}
\end{equation}
For this ellipticity tensor, only the channels represented by Eqs.~\eqref{Qshear.eqn} and~\eqref{QshearEB.eqn} are active.
Aligning the major axis of the ellipse along the crystallographic axes $x$ or $y$ activates the channel in Eq.~\eqref{QshearPlus.eqn}, but turns off the other two channels because the resulting $Q$ tensor will have a zero off-diagonal component $Q_{xy}$. Therefore, aligning the elliptic axis at $45^\circ$ is superior, as it keeps two channels active: it provides both a background ${\cal B}_2$ from Eq.~\eqref{Qshear.eqn} and a field-induced ${\cal B}_3$ from Eq.~\eqref{QshearEB.eqn}. 

On the other hand, we have also found that to maximize the benefit from the nodal line term in Eq.~\eqref{nodalline}, the in-plane magnetic field $\vec B$ must be oriented at $45^\circ$ as well. Combining these two points means that orienting both the dot's elliptical axis and $\vec B$ at $45^\circ$ also allows the random $\vec P$ arising from the Ge distribution to be leveraged, as described by Eq.~\eqref{QshearEB.eqn}. In cases where a deterministic $\vec P$ can be imposed on the quantum dot, the highest coupling strength in Eq.~\eqref{QshearEB.eqn} is achieved when $\vec B$ and the asymmetry vector $\vec P$ are perpendicular.  
To explore this further, note that setting $Q_{xy} = 0.5$ yields ${\cal B}_3 \propto (P_xB_x - P_yB_y)$. A coordinate rotation by $45^\circ$ transforms this into $P'_x B'_y$, revealing a $\sin\theta$ angular dependence, where $\theta$ is the angle between $\vec P$ and $\vec B$. This dependence is maximized when $\vec P$ and $\vec B$ are perpendicular. Furthermore, this $\sin\theta$ behavior can serve as an experimental signature to verify the presence of such terms. For a quantum dot with a fixed $Q$-tensor and fixed $\vec P$, rotating the in-plane magnetic field should produce a $\sin\theta$ variation in the valley splitting.

Interestingly, the topological nodal line term in Eq.~\eqref{nodalline} under an in-plane $\vec B$-field produces a $\sin 2\theta_0$ dependence, arising from the $k_xk_y$ factor, where $\theta_0$ is the angle relative to the crystallographic $x$-axis. In a realistic experiment with sufficiently small quantum dots, both these harmonics—$\sin\theta$ and $\sin 2\theta_0$—are expected to be present in valley splitting maps.

\section{Summary and outlook}

In this work, we have precisely defined the valley degree of freedom by specifying its transformation properties under the group of the wave vector at the $X$ point of the Brillouin zone, from which the valleys originate. 
To emphasize the key distinctions between the transformation properties of the valley degree of freedom and those of spin, we introduce the term \emph{valleyor} to distinguish it from \emph{spinor}. 
By determining the transformation properties of valleyors under the space group and time-reversal symmetry, we were able to identify a composite valley-magnetic field $\vec{\cal B}$ that allows for a Zeeman-like coupling to the valley degree of freedom $\vec\tau$ via $\vec{\cal B}\cdot\vec \tau$.
The possible forms of this coupling are summarized in Tab.~\ref{VMF.tab}. 
Importantly, our scenario includes the possibility of producing a ${\cal B}_3$ term that couples to $\tau_3$. This coupling is absent in known scenarios based on scattering by a scalar potential between the two valleys. This term is capable of producing a non-zero valley splitting in regions where mechanisms coupling only to $(\tau_1,\tau_2)$ fail to split the valleys.

Our analysis has further reproduced a band structure term in Eq.~\eqref{nodalline} that reveals a nodal line in bulk silicon with the precise form vanishing of the band-separation at the node given by Eq.~\eqref{nodalline}. Although this nodal line lies approximately $80$ meV above the Si valley minima, in quantum dots it can provide an additional resource for valley coupling, thereby inducing a nodal-line-mediated mechanism for valley splitting. 
In quantum dots, the ellipticity of a dot—encoded into a two-dimensional rank-two tensor $Q_{ab}$—serves as an additional resource. In terms of its symmetry properties, this tensor behaves similarly to the strain tensor. It provides "shear" components that can, either alone or in combination with the physical strain tensor, generate valley-magnetic field components ${\cal B}_2$ and/or ${\cal B}_3$.

Our present analysis has focused on the single group of the wave vector at the $X$ point, meaning spin degrees of freedom were not included in the transformation properties. Our ongoing work on the double group promises to reveal novel forms of spin-orbit coupling that have no analogues in valley-less systems. This approach allows for the classification of all possible forms of spin-valley coupling~\cite{Jock2022,Cai2023}.

Another direction for extending our analysis is to employ the symmetries of the $\Delta_1$ and $\Delta_{2'}$ bands to construct real-space orbitals. This would enable the design of a symmetry-inspired microscopic model of silicon that focuses specifically on the conduction band and its valley degrees of freedom to address spatial inhomogeneities. Such a model is expected to reveal an interesting interplay between transverse inhomogeneities and valley splitting.

\section{Acknowledgments} We thank Lars Schreiber for insightful discussions. S.A.J. was supported by EinQuantumNRW. S.A.J. made use of DeepSeek AI to improve clarity and flow of the text. 

\bibliography{Refs}



\end{document}